\documentclass[cameraready]{Interspeech}
\usepackage{multirow}
\usepackage{makecell} 

\title{Comparing Self-Supervised and Domain-Invariant Features for Cross-Domain Voice Phishing Detection}

\author[affiliation={1,2}, orcid=0000-0002-4096-9532]{Jeongmin}{Lee}
\author[affiliation={1}]{Seung}{Yun}
\author[affiliation={1}]{Minkyu}{Lee}
\author[affiliation={1}]{Ran}{Han}
\author[affiliation={1,2}]{Yoonkyu}{Woo}
\author[affiliation={1,2}, orcid=0000-0002-2851-5124, correspondingauthor]{Jinxia}{Huang}

\address{
    $^1$ Electronics and Telecommunications Research Institute, Daejeon, Korea \\
    $^2$ University of Science and Technology, Daejeon, Korea
}

\email{\{faraway, syun, mk, ran.han, yoon303, hgh\}@etri.re.kr}


\newcommand{\blue}[1]{#1}

\usepackage{comment}

\begin{document}

\maketitle

\begin{abstract}
Voice phishing detection faces three critical challenges: real criminal recordings are unavailable due to privacy constraints; when available, only a handful of samples exist, insufficient for fine-tuning; and lightweight acoustic-only detection is needed as an alternative to large self-supervised models. We compare domain-invariant prosodic features and self-supervised representations (HuBERT, wav2vec2.0) through cross-domain evaluation---training on scenario-based actor recordings and testing on authentic criminal calls. Domain-invariant prosodic features achieve 69.5\% F1 zero-shot and 71.0\% with 5-shot learning. HuBERT achieves highest performance (94.2\% F1, 5-shot), while wav2vec2.0 exhibits a precision-oriented detection profile (90.2\% F1 with 99.4\% precision, 5-shot). These findings reveal fundamental trade-offs: domain-invariant features enable zero-shot deployment when no real data exists, while SSL methods achieve higher performance but require real samples and compute.
\end{abstract}

\section{Introduction}

\blue{Voice phishing (vishing) causes substantial financial losses worldwide. Real-time call monitoring is predominantly constrained to on-device processing: transmitting raw audio to external servers raises privacy and security concerns, preventing centralized data collection in most investigative scenarios.} \blue{Detection systems therefore face severe constraints in practice—real criminal recordings are difficult to access for training, and even when accessible, only a handful of samples exist, insufficient for fine-tuning models.} These constraints necessitate cross-domain approaches that train on accessible or constructable scenario-based data and generalize to authentic criminal calls.

\blue{Self-supervised learning (SSL) models such as HuBERT and wav2vec2.0 have achieved strong results on diverse speech tasks via large-scale pretraining.} \blue{However, their applicability to voice phishing detection remains unclear: do SSL models generalize across the domain shift from controlled scenario-based recordings---where actors read fraud scripts in studio environments constructed for training---to authentic criminal calls?} Alternatively, domain-invariant prosodic features selected for cross-domain stability may offer different trade-offs in terms of performance and computational requirements. However, systematic evaluation comparing these approaches across varying data availability scenarios has not been conducted.

We present a systematic comparison of domain-invariant prosodic features and self-supervised representations (HuBERT, wav2vec2.0) for cross-domain voice phishing detection. We evaluate these approaches in a simulated-to-authentic cross-domain setting, where models trained on scenario-based data are tested on real criminal calls. Through zero-shot, 1-shot, and 5-shot evaluation, we reveal fundamental trade-offs: while HuBERT achieves the highest performance (94.2\% F1, 5-shot), domain-invariant prosodic features demonstrate competitive zero-shot performance (69.5\% F1) without requiring any real criminal samples.

Our contributions are three-fold: (1) We present the first systematic comparison of domain-invariant prosodic features and SSL models for cross-domain voice phishing detection on authentic criminal recordings. (2) We analyze performance across zero-shot and few-shot settings, revealing fundamental trade-offs between accuracy, computational cost, and sample efficiency. (3) We provide practical deployment guidelines: domain-invariant features suit resource-constrained or cold-start scenarios (69.5\% zero-shot), while SSL methods excel when target-domain samples and computational resources are available (94.2\% with 5-shot).

\section{Related Work}

Voice phishing detection has received growing attention, with recent work exploring multimodal approaches combining text and audio~\cite{kim2025multimodal} and fine-tuning language models on transcripts~\cite{sim2025finetuning}. However, these approaches require substantial computational resources for text extraction and language model deployment. We focus on acoustic-only detection that can operate under resource constraints, comparing domain-invariant prosodic features with self-supervised representations for deployability and generalization.

Self-supervised learning models such as wav2vec 2.0~\cite{baevski2020wav2vec2} and HuBERT~\cite{hsu2021hubert} have achieved strong performance on diverse speech tasks including ASR, speaker verification, and emotion recognition through large-scale pretraining~\cite{wang2021finetuned}. While cross-domain robustness of SSL models has been demonstrated in ASR~\cite{zhu2023boosting}, their effectiveness for cross-domain voice phishing detection remains unexplored---specifically, whether they can generalize from scenario-based actor recordings to authentic criminal calls without requiring substantial target-domain data.

Prior work has demonstrated that prosodic features can effectively detect deceptive speech in controlled settings. Levitan et al.~\cite{levitan2016combining} show that acoustic-prosodic features including F0, energy, and duration patterns improve deception detection accuracy. Chen et al.~\cite{chen2020acoustic} identify specific prosodic cues that predict both deception and perceived trustworthiness, while Goupil et al.~\cite{goupil2021common} establish prosodic signatures associated with certainty and honesty perception. However, these studies validate prosodic features within single controlled domains. Moreover, authentic criminal recordings are often unavailable for training in real-world deployments due to privacy regulations, necessitating approaches that can generalize without extensive target-domain access.

Cross-domain adaptation methods typically require substantial target-domain data for model fine-tuning or adversarial training. For example, Zhu et al.~\cite{zhu2023boosting} combine self-supervised pretraining with pseudo-labeling on target data, while Meng et al.~\cite{meng2019attentive} use adversarial training across domains. However, in voice phishing detection, only limited authentic criminal recordings may be accessible. We address this constraint by separating feature selection from classifier training: we use Cohen's $d$---a standardized measure of effect size between two
  distributions across different conditions~\cite{cohen1988statistical}---to identify prosodic features with small distributional differences ($d < 0.5$) between domains. Features meeting this criterion exhibit reduced sensitivity to domain-specific factors, enabling effective classifier training on source data with minimal target samples required only for feature selection. We evaluate this approach across scenarios with varying target-domain availability (0, 1, 5-shot).

\section{Methods}

\subsection{Dataset Construction}

To study cross-domain generalization in voice phishing detection, we construct a Korean speech corpus comprising four datasets arranged along two axes: class (VP vs.\ non-VP) and domain (scenario-based vs.\ authentic real-world). All audio is standardized to 8 kHz mono to reflect telephony bandwidth conditions. Speaker-independent splits are applied across all partitions to prevent speaker overlap between training and test sets.

\textbf{Voice Phishing Datasets.}
\textit{Scenario-based VP} consists of 406 utterances recorded by professional voice actors reading scripted voice phishing scenarios in controlled studio environments. This dataset was constructed for this study.

\textit{Authentic criminal VP} consists of 456 utterances extracted from real telephone scam cases, sourced from a national financial supervisory authority~\cite{fssdata}. Of these, 406 utterances form the evaluation set; for few-shot conditions, $k \in \{0, 1, 5\}$ support examples are randomly sampled from a separate 50-utterance holdout, strictly disjoint from the evaluation set.

\textbf{Non-Voice Phishing Datasets.}
\textit{Scenario-based financial consultation} is drawn from an AI Hub financial domain corpus~\cite{aihubdata_scenario}, comprising simulated customer--agent telephone interactions at 8 kHz. We use 406 utterances for training and 203 for evaluation.

\textit{Authentic telephone consultation} is drawn from the AI Hub low-bandwidth telephony speech corpus~\cite{aihubdata_real}, which covers four service domains: public services (31\%), e-commerce (35\%), education (23\%), and HR (11\%). We proportionally sample 203 utterances to preserve the original domain distribution. Table~\ref{tab:dataset} summarizes the dataset composition.

\begin{table}[t]
  \caption{Dataset composition for cross-domain voice phishing detection. For few-shot adaptation, $k \in \{0, 1, 5\}$ support examples are randomly sampled from a 50-utterance authentic criminal VP set, strictly disjoint from the test set. The test set remains fixed at 812 utterances across all conditions. Train/test partitions are sample-disjoint.
}
  \label{tab:dataset}
  \centering
  \footnotesize
  \setlength{\tabcolsep}{3pt}
  \begin{tabular}{l l r r r r}
    \toprule
    \textbf{Dataset} & \textbf{Class} & \textbf{Train} & \textbf{Test} & \blue{\textbf{Avg (s)}} & \blue{\textbf{Total (h)}} \\
    \midrule
    \multicolumn{6}{l}{\textit{Source domain (scenario-based)}} \\
    \quad Scenario VP              & VP     & 406 & ---  & \blue{506} & \blue{57.1} \\
    \quad Scenario consult.        & Non-VP & 406 & 203 & \blue{220} & \blue{37.1} \\
    \midrule
    \multicolumn{6}{l}{\textit{Target domain (authentic)}} \\
    \quad Authentic VP             & VP     & --- & 406 & \blue{327} & \blue{36.9} \\
    \quad Authentic consult.       & Non-VP & --- & 203 & \blue{104} & \blue{5.9} \\
    \midrule
    \multicolumn{2}{l}{\textbf{Total}} & \textbf{812} & \textbf{812} & \blue{---} & \blue{\textbf{137.0}} \\
    \bottomrule
  \end{tabular}
\end{table}

The cross-domain evaluation trains classifiers on source-domain data and tests on target-domain data. As detailed in Section~\ref{sec:features}, domain-invariant feature selection is performed as a preliminary offline analysis using authentic VP samples; the classifier is trained exclusively on scenario-based data, with few-shot settings incorporating samples drawn from the holdout pool into training.

\subsection{Feature Extraction}
\label{sec:features}

\textbf{Domain-Invariant Prosodic Features.}
Domain-invariant feature selection is conducted as a preliminary offline analysis, separate from classifier training: authentic VP samples identify which acoustic features remain stable across the scenario-to-authentic domain shift, and the selected feature set is then fixed for all subsequent classifier training.

\begin{figure}[!htbp]
  \centering
  \includegraphics[width=1.05\columnwidth]{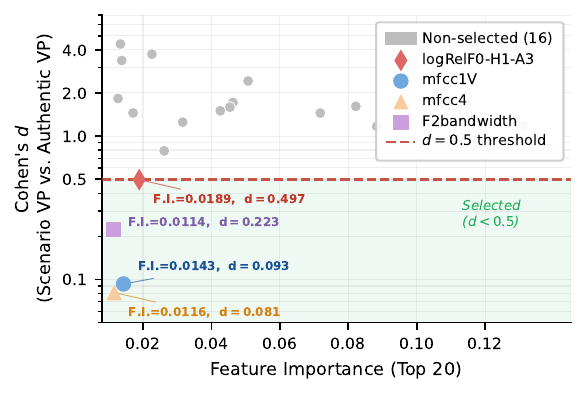}
  \caption{RF feature importance vs.\ Cohen's $d$ for the Top-20 eGeMAPS candidates. Dashed line: $d{=}0.5$ selection threshold; shaded region: selected zone ($d < 0.5$). Four selected features are annotated with RF importance (F.I.) and Cohen's $d$.}
  \label{fig:feature_selection}
\end{figure}

We extract eGeMAPS v02~\cite{eyben2016gemaps} features (88 functionals) using openSMILE 3.0~\cite{eyben2010opensmile}. Domain-invariant features are identified via a two-stage pipeline: (1) a Random Forest classifier trained on authentic criminal VP vs. scenario-based non-VP retains the top 20 features by importance, targeting class-boundary discriminability; (2) Cohen's $d$~\cite{cohen1988statistical} between scenario-based and authentic VP retains features with $d < 0.5$ (small-to-moderate effect size), targeting domain-boundary stability. Figure~\ref{fig:feature_selection} plots RF importance against Cohen's $d$ for all Top-20 candidates, with the four selected features annotated.

\blue{Four features satisfy this criterion: logRelF0-H1-A3, mfcc1V, mfcc4, and F2bandwidth. logRelF0-H1-A3 reflects spectral tilt under vocal stress~\cite{sigmund2012influence,eyben2016gemaps}; mfcc1V and mfcc4 capture spectral envelopes tied to voice quality and articulation in deceptive speech~\cite{levitan2016combining,chen2020acoustic}; and F2bandwidth indexes formant bandwidth shifts under articulatory tension~\cite{roy2009articulatory}, consistent with prosodic markers of perceived honesty~\cite{goupil2021common}.}

\clearpage

\begin{table}[h]
  \centering
  \caption{Cross-domain results (LR classifier, \%). $k$: number of authentic VP support samples. Bold: best per metric per $k$ condition.}
  \label{tab:results}
  \begin{tabular}{l l c r r r}
    \toprule
    \textbf{Method} & \textbf{Features} & $\boldsymbol{k}$\textbf{-shot} & \textbf{F1} & \textbf{Rec.} & \textbf{Prec.} \\
    \midrule
    \multirow{3}{*}{Prosodic} & \multirow{3}{*}{\makecell[l]{88\\(baseline)}}
      & 0 &  3.8 &  2.0 & 66.7 \\
    & & 1 &  8.8 &  4.7 & 73.1 \\
    & & 5 & 85.2 & 90.4 & 80.5 \\
    \midrule
    \multirow{3}{*}{Prosodic} & \multirow{3}{*}{\makecell[l]{4\\(selected)}}
      & 0 & \textbf{69.5} & \textbf{79.1} & 62.0 \\
    & & 1 & 68.8 & 76.8 & 62.3 \\
    & & 5 & 71.0 & 81.0 & 63.1 \\
    \midrule
    \multirow{3}{*}{HuBERT} & \multirow{3}{*}{768}
      & 0 & 58.3 & 41.4 & 98.8 \\
    & & 1 & \textbf{88.8} & \textbf{84.2} & 94.0 \\
    & & 5 & \textbf{94.2} & \textbf{99.3} & 89.6 \\
    \midrule
    \multirow{3}{*}{wav2vec2.0} & \multirow{3}{*}{768}
      & 0 & 36.2 & 22.2 & \textbf{98.9} \\
    & & 1 & 46.0 & 30.0 & \textbf{98.4} \\
    & & 5 & 90.2 & 82.5 & \textbf{99.4} \\
    \bottomrule
  \end{tabular}
\end{table}

\textbf{Self-Supervised Feature Extraction.}
For SSL-based feature extraction, we use HuBERT-Base~\cite{hsu2021hubert}\footnote{\url{https://hf.co/facebook/hubert-base-ls960}} and wav2vec2.0-Base~\cite{baevski2020wav2vec2}\footnote{\url{https://hf.co/facebook/wav2vec2-base-960h}}, both pretrained on LibriSpeech 960h (12 transformer layers, 768-dimensional hidden states, $\sim$94M parameters each). Frame-level representations from the final hidden layer are mean-pooled to obtain a 768-dimensional utterance embedding per sample. SSL encoder weights are kept frozen; no fine-tuning is performed on any domain-specific data.

\subsection{Classification Setup}

We use Logistic Regression (LR; L2 regularization, $C{=}1.0$) as the classifier for all feature types. LR is chosen for its superior stability over alternative classifiers in few-shot conditions, computational efficiency enabling on-device deployment, and well-calibrated linear decision boundary on normalized features. Features are standardized using per-feature mean and variance computed from training data only.

For each condition, $k \in \{0, 1, 5\}$ authentic VP samples from the holdout pool are appended to the fixed scenario-based training set; the test set remains constant at 812 utterances across all conditions (see Table~\ref{tab:dataset}).

\subsection{Evaluation Protocol}

We report F1 score, Recall (true positive rate for VP class, critical for security applications) and Precision (false alarm rate on legitimate calls; affects user experience in real deployments). The cross-domain evaluation measures generalization from scenario-based training to authentic real-world test data.

\blue{In this work, zero-shot denotes classifier transfer without authentic VP samples in classifier training. The preliminary Cohen's $d$ analysis (Section~\ref{sec:features}) identifies domain-stable acoustic features from distributional statistics alone; decision boundaries are learned exclusively from scenario-based data.}

Few-shot settings ($k \in \{0, 1, 5\}$ authentic VP samples drawn from holdout) evaluate performance under varying target-domain data availability, with holdout and test sets strictly disjoint to prevent data leakage.

\section{Results and Discussion}

\begin{figure}[b]
  \centering
  \includegraphics[width=\columnwidth]{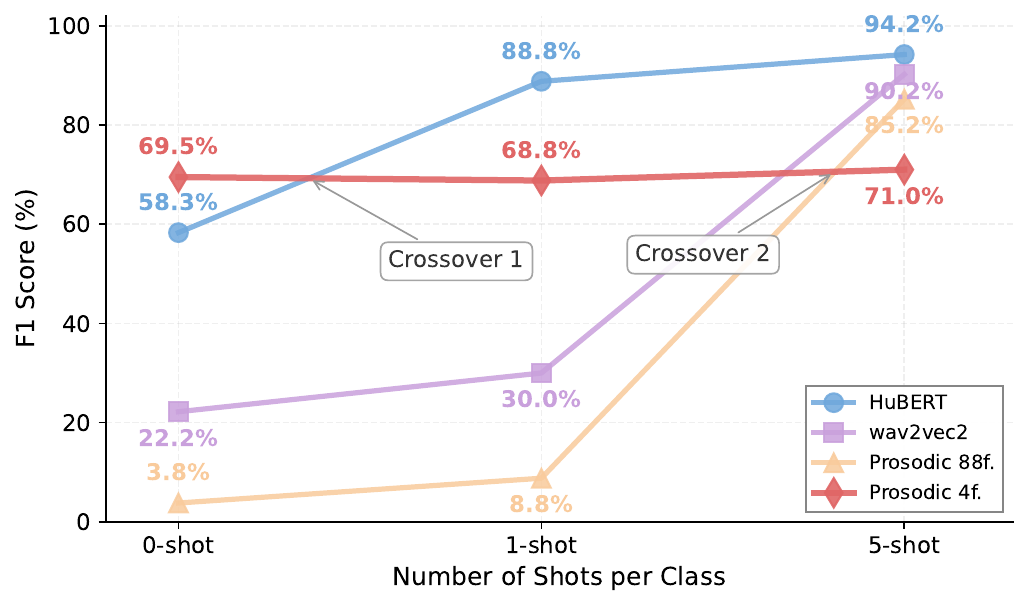}
  \caption{F1 score vs.\ number of authentic VP samples (0, 1, 5-shot). Crossover~1: regime transition where HuBERT surpasses domain-invariant Prosodic~4f. Crossover~2: full Prosodic~88f. overtakes Prosodic~4f. with 5-shot adaptation.}
  \label{fig:curves}
\end{figure}

Table~\ref{tab:results} reports F1, VP Recall and Precision across all conditions. Figure~\ref{fig:curves} visualizes F1 learning curves as a function of target-domain support samples.

\textbf{Zero-shot deployment capability.}
At zero-shot, domain-invariant prosodic features (4f.) achieve 69.5\% F1 without any authentic criminal samples. \blue{The full 88-feature set---lacking domain-invariant filtering---collapses to 3.8\%, a 65.7-point gap confirming Cohen's $d$ filtering is a prerequisite, not optional. Among all methods, 4f. also leads at zero-shot over SSL: HuBERT reaches 58.3\% and wav2vec2.0 36.2\%, as the scenario-to-authentic shift degrades SSL representations without target-domain calibration.} \blue{This 4f. advantage persists until Crossover~1 ($k{\approx}0.4$, Figure~\ref{fig:curves}), beyond which SSL representations scale rapidly with target-domain samples.}

\textbf{Few-shot scaling and the 88f. collapse.}
\blue{With 5-shot, HuBERT reaches 94.2\% F1 (+35.9~pp over zero-shot), the strongest result when even a few authentic samples are available.} wav2vec2.0 shows a similarly steep gain (36.2\% $\rightarrow$ 90.2\%). By contrast, domain-invariant features (4f.) show only marginal improvement (69.5\% $\rightarrow$ 71.0\%), reflecting that the selected feature set is already near its performance ceiling with scenario-based training alone. The full 88-feature set reveals the necessity of domain-invariant filtering: without it, prosodic features collapse at zero- and one-shot (3.8\%, 8.8\%), recovering only at 5-shot (85.2\%), at which point the 88f. set undergoes a regime transition (Crossover~2, Figure~\ref{fig:curves}), surpassing the domain-invariant 4f. subset. This confirms that the full eGeMAPS set is dominated by domain-sensitive dimensions that transfer poorly from studio to telephone conditions.

\textbf{Precision--Recall trade-off across methods.}
The methods exhibit distinct precision--recall profiles with practical implications. HuBERT at 5-shot achieves near-perfect Recall (99.3\%), missing only 3 of 406 criminal calls, but at the cost of reduced precision (89.6\%), generating false alarms on legitimate calls. wav2vec2.0 at 5-shot shows the inverse pattern: Recall of 82.5\% with near-perfect precision (99.4\%), producing almost no false alarms. This distinction matters in real deployment: HuBERT prioritizes detection completeness (critical for security), while wav2vec2.0 minimizes false alarms (critical for service quality). \blue{Domain-invariant prosodic features lie between these extremes across all shot conditions (Precision $\approx$ 62--63\%, Recall $\approx$ 77--81\%) without target-domain samples.}

\textbf{HuBERT vs.\ wav2vec2.0 at zero-shot.}
Although HuBERT and wav2vec2.0 share identical architecture (12 transformer layers, 768-dim, LibriSpeech 960h pretraining), their zero-shot generalization differs markedly: HuBERT achieves 58.3\% F1 versus 36.2\% for wav2vec2.0. Both methods exhibit high precision at zero-shot (98.8\% and 98.9\%, respectively) but differ substantially in recall (41.4\% vs.\ 22.2\%), indicating that wav2vec2.0 representations require more target-domain adaptation before reliably activating on authentic criminal VP. This gap may reflect pretraining objective differences: HuBERT's cluster-prediction objective may produce representations that more directly encode phonetic and prosodic contrasts relevant to voice phishing, while wav2vec2.0's contrastive objective may favour distinctiveness across nearby frames rather than long-range prosodic patterns.

\textbf{Practical deployment guidelines.}
These results suggest a deployment decision framework based on target-domain data availability. When no authentic criminal VP samples are available (cold-start), domain-invariant prosodic features (4f.) provide reliable 69.5\% F1 detection with minimal computational overhead. When even 5 authentic samples can be obtained through legal proceedings, HuBERT achieves 94.2\% F1 with near-perfect recall, prioritizing detection completeness; wav2vec2.0 achieves 90.2\% F1 with near-perfect precision, prioritizing service quality. The full 88-feature prosodic set is unsuitable for cold-start scenarios but becomes competitive (85.2\%) with 5-shot adaptation.

\textbf{Ablation: Individual feature contributions within the 4-feature set.}
We evaluate all 15 subsets (4 individual, 6 pairs, 4 triples, 1 full) of the domain-invariant features under zero-shot, 1-shot, and 5-shot cross-domain protocols to assess the individual contribution of each feature to cross-domain detection performance. Table~\ref{tab:ablation} reports the primary comparison between mfcc1V alone and the full 4-feature set across shot conditions; all subsets containing mfcc1V exceed 4f in F1, while all subsets without mfcc1V fall below. We additionally report Sensitivity (VP true positive rate) and Specificity (non-VP true negative rate) to characterise the detection trade-off within this subset.

mfcc1V consistently achieves the highest F1 and Sensitivity across all shot conditions, confirming its role as the primary VP detection driver. The full 4-feature set shows lower Sensitivity---reflecting a more conservative decision boundary---but improved Specificity at 1-shot (+3.2\,pp) and 5-shot (+2.7\,pp), indicating reduced false-alarm rates on genuine calls when minimal target-domain data is available. This reveals a detection trade-off: mfcc1V maximises the VP capture rate, while 4f offers complementary false-alarm suppression under realistic few-shot deployment conditions.

\begin{table}[h]
  \centering
  \caption{mfcc1V vs.\ 4 features (selected) (LR classifier, \%). Sens. = VP correct detection rate (Sensitivity); Spec. = Specificity on within- and out-of-distribution Non-VP combined; Spec. (OOD) = Specificity on out-of-distribution Non-VP only (authentic telephone consultation, $n$=203). Bold: best per metric per condition.}
  \label{tab:ablation}
  \small
  \setlength{\tabcolsep}{4pt}
  \begin{tabular}{l r r r r r r}
    \toprule
    & \multicolumn{3}{c}{\textbf{mfcc1V}} & \multicolumn{3}{c}{\textbf{4 features (selected)}} \\
    \cmidrule(lr){2-4} \cmidrule(lr){5-7}
    \makecell[l]{$\boldsymbol{k}$\textbf{-shot}\\~} & \makecell{Sens.\\~} & \makecell{Spec.\\~} & \makecell{Spec.\\(OOD)} & \makecell{Sens.\\~} & \makecell{Spec.\\~} & \makecell{Spec.\\(OOD)} \\
    \midrule
    0 & \textbf{87.9} & \textbf{52.0} & 38.4 & 79.1 & 51.5 & \textbf{39.9} \\
    1 & \textbf{88.2} & 51.7 & 37.9 & 75.9 & \textbf{54.9} & \textbf{38.9} \\
    5 & \textbf{87.7} & 52.0 & 37.4 & 76.8 & \textbf{54.7} & \textbf{37.9} \\
    \bottomrule
  \end{tabular}
\end{table}

The mixed Non-VP test set contains a within-distribution component (scenario-based consultation, same scripted domain as training Non-VP) that partially masks this Specificity pattern: at zero-shot in the full evaluation set, mfcc1V Specificity (52.0\%) marginally exceeds 4f (51.5\%), rendering the complementary advantage ambiguous. Evaluating on the authentic Non-VP subset only (Spec.\ (OOD), $n$=203) removes this confound---the 4f Specificity advantage over mfcc1V becomes consistent across all shot conditions ($k$=0: 39.9\% vs.\ 38.4\%; $k$=1: 38.9\% vs.\ 37.9\%; $k$=5: 37.9\% vs.\ 37.4\%), confirming that the complementary false-alarm suppression of the full feature set is more salient under purely out-of-distribution evaluation.

Among the remaining 13 subsets, logRelF0-H1-A3---which carries the highest domain-discriminativeness ($d{=}0.497$)---yields near-chance F1 (18.8\%) individually. This reveals a key distinction: high Cohen's $d$ reflects large domain shift, not discriminative utility, and its inclusion in 4f is the primary source of Sensitivity reduction.

The full 4-feature set is retained as the primary system for two reasons. First, 4f demonstrates superior Specificity under few-shot conditions — the deployment scenario most relevant to real-world telephony systems — providing improved false-alarm suppression precisely when minimal target-domain calibration data is available. Second, the feature set was fixed prior to experiments based on the domain-invariance criterion ($d < 0.5$); post-hoc selection of mfcc1V from ablation results would constitute test-set optimisation, as the ablation is designed to characterise individual contributions rather than re-optimise the feature set.

\vspace{-0.35em}

\section{Conclusion}

We presented a systematic cross-domain evaluation of domain-invariant prosodic features and self-supervised representations (HuBERT, wav2vec2.0) for voice phishing detection, training on scenario-based recordings and testing on authentic criminal calls. Our results reveal a clear regime structure governed by target-domain data availability. At zero-shot, domain-invariant prosodic features (4f.) achieve 69.5\% F1---outperforming both SSL models---while the unfiltered 88-feature set collapses to 3.8\%, confirming that Cohen's $d$-based filtering is a prerequisite for cold-start deployment. With 5-shot adaptation, SSL models surpass prosodic features: HuBERT reaches 94.2\% F1 with near-perfect recall, and wav2vec2.0 achieves 90.2\% with near-perfect precision. \blue{Ablation identifies mfcc1V as the primary detection driver: domain stability alone does not imply discriminative utility. Whether this regime structure extends beyond Korean remains an open question for cross-lingual validation.}

\blue{From a deployment perspective, the two approaches are complementary. Domain-invariant prosodic features need no pretrained model and run on CPU, suiting real-time on-device monitoring without GPU. SSL models ($\sim$94M parameters each) require GPU but deliver higher accuracy once minimal labeled data is available. The choice depends on hardware constraints.}

\vspace{-0.35em}

\section{Acknowledgments}
This work was supported by the Institute of Information and Communications Technology Planning and Evaluation (IITP) grants funded by the Korea government (MSIT) (No. RS-2025-02215393, Development of Detection and Prediction Technology for New and Unknown Voice Phishing; No. RS-2019-II190004, Development of Semi-supervised Learning Language Intelligence Technology and Korean Tutoring Service for Foreigners).

\section{Generative AI Use Disclosure}
Generative AI tools (Claude, Anthropic) were used to assist with code development and manuscript language editing. All AI-generated code was reviewed, tested, and validated by the authors before use. AI tools were not used for research design, hypothesis generation, experimental methodology, data analysis, result interpretation, or scientific conclusions. Any AI-assisted manuscript revisions were limited to language polishing and clarity improvements, while all scientific content was written and verified by the authors.

\bibliographystyle{IEEEtran}
\bibliography{mybib}

\end{document}